\documentstyle[11pt,aaspp4]{article}
\begin{document}
\title{Yet Another Model of Gamma-Ray Bursts}
\author{J. I. Katz \\ katz@wuphys.wustl.edu}
\notetoeditor{Until June 1, 1997 address correspondence to author c/o
T. Piran, Racah Institute of Physics, Hebrew University, Jerusalem 91904,
Israel or jonathan@shemesh.fiz.huji.ac.il.  After June 1, 1997 address
correspondence to author at Department of Physics, Washington University,
St. Louis, Mo. 63130 or katz@wuphys.wustl.edu.}
\affil{Racah Institute of Physics, Hebrew University, Jerusalem 91904, 
Israel \\ and \\ Department of Physics and McDonnell Center for the Space
Sciences \\ Washington University, St. Louis, Mo. 63130}
\authoremail{katz@wuphys.wustl.edu}
\begin{abstract}
\cite{SP97a} have demonstrated that the time structure of 
gamma-ray bursts must reflect the time structure of their energy release.  A
model which satisfies this condition uses the electrodynamic emission of 
energy by the magnetized rotating ring of dense matter left by neutron star 
coalescence; GRB are essentially fast, high field, differentially rotating
pulsars.  The energy densities are large enough that the power appears as an
outflowing equilibrium pair plasma, which produces the burst by baryon 
entrainment and subsequent internal shocks.  I estimate the magnetic field
and characteristic time scale for its rearrangement, which determines the 
observed time structure of the burst.  There may be quasi-periodic
oscillations at the rotational frequencies, which are predicted to range up
to 5770 Hz (in a local frame).  This model is one of a general class of 
electrodynamic accretion models which includes the \cite{B76} and 
\cite{L76} model of AGN, and which can also be applied to black hole 
X-ray sources of stellar mass.  The apparent efficiency of nonthermal 
particle acceleration is predicted to be 10--50\%, but higher values are 
possible if the underlying accretion flow is super-Eddington.  Applications
to high energy gamma-ray observations of AGN are briefly discussed.
\end{abstract}
\section{Introduction}
More than a hundred models of gamma-ray bursts (GRBs), including soft
gamma repeaters (SGRs), have been published (\cite{N94}), yet there is
at present no satisfactory and generally accepted model of GRBs.  Even
early data hinted at cosmological distances (\cite{UC75,vdB83}).
At these distances the inferred energy release of $\sim 10^{51}$ erg in a 
region constrained by their shortest observed times of variation implies 
(\cite{CR78}) the creation of an equilibrium pair and photon fireball
with temperature $> 10^{10\,\circ}$K.  

The expected properties of radiation-pair fireballs (\cite{G86})
do not resemble the observed properties of GRB.  Fireballs are expected to 
produce a brief ($< 1$ ms if their energy is released in a region of neutron
star dimensions) pulse of black-body radiation, very different from the 
properties of observed (\cite{F94}) GRB, which have complex multipeaked time
structure with durations in the range $10^{-2}$--$10^3$ s and nonthermal 
spectra.

For these reasons the hypotheses of cosmological distances and, by 
implication, of fireball models, were not generally accepted until data
from BATSE made the geometrical case for them compelling (\cite{M92,P92}).
This left the problem of reconciling the observed properties of GRB
with the predicted properties of fireballs.  \cite{SP90} had
demonstrated that even a small contamination of the fireball with
baryonic matter (and its associated non-annihilating electron excess) leads
to the conversion of nearly all the fireball energy to the kinetic energy
of a relativistic shell of baryons.  \cite{RM92} then 
pointed out that this shell could form a collisionless shock when it
encountered surrounding diffuse matter, and suggested that radiation by
the shocked matter could resolve the discrepancy between the predicted
and observed properties.  Relativistic shocks plausibly produce a
nonthermal spectrum, optically thin at low frequencies (thus avoiding
a Wien slope) and with power law extensions to higher frequencies.  The 
complex temporal structure of GRB was attributed to the known complex
spatial structure of diffuse matter, such as the interstellar medium.
This suggestion was widely accepted (with a sigh of relief), and led to
a great deal of work on the properties of fireballs and relativistic 
shocks.

Most fireball models of GRB have suggested that they are produced by the 
coalescence (\cite{P86,E89}) or formation (\cite{D92}) of neutron stars.  A
fraction of the binding energy escapes as neutrinos and forms
$e^\pm$ pairs by interactions between two neutrinos.  These events release 
energy over several seconds, with a smooth time profile, as a consequence of
the diffusion of neutrinos through matter dense enough to be opaque even to
them (as demonstrated empirically by the duration of neutrino emission
from SN 1987A).  \cite{KC96} pointed out that such a model could
not explain short GRB, with durations $< 2$ s, and suggested collisions
between two neutron stars, from which the escape of the neutrinos would
be accelerated by the expansion of the debris.  This argument is also
applicable to the fine temporal substructure of long GRB, thus excluding
processes depending on neutrino diffusion as the explanation of almost
any GRB.  The most popular version of this model involves the coalescence 
of two neutron stars, which has also been calculated (\cite{D94,JR96,Ma96})
to be an inadequate source of neutrinos because coalescence is nearly 
adiabatic (the subsonic velocity of convergence does not make shocks) and 
does not sufficiently heat the cold degenerate neutron star matter.

\cite{SP97a} demonstrated, given quite general assumptions, that
external shocks can only produce smooth, single peaked GRB, quite unlike
observed GRB.  This conclusion is very general because it is
essentially geometrical; it does not depend on arguments concerning
neutrino diffusion times or other microscopic physical processes.  It is
applicable to any single release of energy, however thin the relativistic
shell and however complex the distribution of surrounding matter, and
therefore excludes even the brief ($\sim 5$ ms) neutrino emission of 
colliding neutron stars as the explanation of GRB.  They showed that a
complex GRB pulse profile can only be the consequence of a complex
history of energy release by the engine which powers the GRB.  This excludes
models in which the GRB energy is released in a single brief event,
whether the coalescence, birth or collision of compact objects.  It implies
than models involving a continuous or intermittent wind (\cite{P86}) are 
more likely to be correct that those involving a single fireball
(\cite{G86}).

This paper proposes a resolution of these difficulties.  As in much
previous work, I begin with coalescing neutron stars which release an 
enormous ($\sim 3 \times 10^{53}$ ergs) amount of energy and which are 
estimated (\cite{NPS91,P91}) to occur at a rate consistent with the rate of 
GRB.  Such a coalescence results in a rapidly differentially rotating object
containing the mass of the two neutron stars.  It is largely supported by 
internal pressure near its axis but more and more by angular momentum at 
larger distances from the axis.  The central core may promptly collapse to a
black hole, or may remain for an extended time as a rapidly differentially
rotating neutron star, depending on the (uncertain) equation of state.  It 
is surrounded by $\sim 0.1\ M_\odot$ of matter, at densities up to $\sim 
10^{14}$ g/cm$^3$ but decreasing outwards, which is largely supported by 
angular momentum and which may be qualitatively described as an equatorial 
bulge or a thick accretion disc.  For convenience and because this term is 
universally used I will refer to it as a disc, although it is unlikely to be
strongly flattened.  This configuration was described by 
\cite{NPP92,U92,U94,T94}.  It is necessary to explain how it can produce the
observed properties of GRB.

I propose that this rotating object produces a GRB by the electrodynamic
processes which turn rotational energy into particle acceleration in pulsars.
The required magnetic energy is only a tiny fraction of the GRB energy, and
is much less than that required in order to explain GRB as the result of
magnetic reconnection.  In \S2 I describe how these processes may work for 
parameters appropriate to GRB.  In \S3 I discuss the origin and estimate the
magnitude of the magnetic field.  The duration, rapidity of variation, and 
temporal complexity of GRB are discussed in \S4, and possible quasi-periodic
oscillations are predicted.  I compare to magnetic field reconnection in 
\S5.  \S6 considers GRB as members of a unified class of electrodynamically 
accreting objects which includes AGN and Galactic black hole X-ray sources 
(BHXS) such as Cyg X-1.  \S7 contains a summary and conclusions.
\section{Electrodynamics}
A rotating magnetized object is a source of energy.  If it has a magnetic 
dipole moment misaligned with its rotation axis it radiates a power given
by the classical magnetic dipole radiation formula.  \cite{GJ69} showed that
even an aligned rotor produces a similar power (in a relativistic wind) if
the surrounding space is filled with plasma.  In the present case the object
is differentially rotating, but that makes no essential difference; 
\cite{B76} and \cite{L76} similarly applied the theory to differentially 
rotating accretion discs in order to explain AGN.
For a large scale ordered magnetic field $B$ a power
$$P_{rw} \sim {B^2 r^6 \Omega^4 \over 2 c^3} \eqno(1)$$
flows outward in a relativistic wind, where $\Omega$ is the (approximately
Keplerian) angular velocity and all parameters are roughly defined 
mass-weighted means over the disc.  Only a fraction $\sim r \Omega/c$ of 
the magnetic field lines are open, so the power density on those field
lines at the surface
$$S_{rw} \sim {B^2 \over 8 \pi} c \left({r \Omega \over c}\right)^3.
\eqno(2)$$
Near the surface of a neutron star, or near the last stable circular
orbit of a black hole, $r \Omega / c \sim 0.5$, so that $S_{rw}$ is
roughly an order of magnitude less than $c$ times the magnetostatic
energy density.  Quite apart from uncertainties in the physical parameters,
a nonrelativistic treatment is only approximate.  For
numerical evaluations I will take $(r \Omega / c)^3 = 0.1$.

The usual estimate of the power of a GRB at cosmological distances is
$P \sim 10^{51}$ erg/s, although this is known at best to order of magnitude
because their distances must be estimated from uncertain statistical
arguments.  Taking a radiating surface area of $10^{13}$ cm$^2$ (a neutron
star or the inner disc around a black hole of a few $M_\odot$) implies
$S \sim 10^{38}$ erg/cm$^2$s.  If $S = S_{rw}$ then $B \sim 10^{15}$ gauss.
This is much larger than known neutron star magnetic fields, but \cite{T94}
discussed the possibility that some neutron stars may be born with such 
fields (such neutron stars would be nearly unobservable as radio pulsars 
because of their rapid spindown).  In \S3 I discuss further the generation 
of magnetic fields of this magnitude as a consequence of neutron star 
coalescence.

In conventional pulsar models the magnetic field is accompanied (in an
inertial frame) by an electric field $O(B \Omega r/c)$ (\cite{GJ69})
which is large enough to pull charge from the stellar surface or to 
produce $e^\pm$ pairs by a process of vacuum sparking (\cite{S71,RS75}).
In the GRB problem the electric fields are even larger than in pulsars 
(because of both the assumed larger $B$ and the large Keplerian $\Omega$ in 
the inner disc), so there is no difficulty in providing the pair plasma 
which is necessary for the aligned rotor to produce power.  

The power density is extremely large; $S_{rw} \sim 10^{38}$ erg/cm$^2$s 
corresponds to an effective temperature $k_B T_e \approx 3$ MeV.  If $k_B 
T_e > 20$ KeV, corresponding to the much lower equilibrium energy
density $E_{eq} \equiv a T_e^4 > 10^{19}$ erg/cm$^3$ and radiated power 
density $S_{eq} \equiv \sigma_{SB} T_e^4 > 10^{29}$ erg/cm$^2$s, the 
energy density is sufficiently large that an opaque equilibrium pair plasma
can form (the temperature is found from the condition that the equilibrium 
pair density has a Thomson scattering mean free path $\sim 1$ km, but is 
very insensitive to this assumed length).  The laws of thermodynamics 
permit this energy and power density to assume the form of a smaller number
of more energetic particles, but interactions among them and with the 
magnetic field will generally lead to rapid equilibration.  This is 
particularly true at higher energy densities, at which the density of 
interacting particles is higher; the energy density required in GRB is 
$\sim 10^9$ times the minimum required for an equilibrium pair plasma, so it
is safe to assume that it forms.  The opaque pair plasma takes the place of 
the dilute transparent fluid of energetic particles found in radio pulsars, 
for which the power density is much smaller ($S \sim 10^{27}$ erg/cm$^2$s 
for the Crab pulsar, and less for others).  Isolated neutron stars with 
sufficiently strong fields and rapid rotation that their power density 
exceeds $S_{eq}$ (such as a newly born neutron star with a field like that
of the Crab pulsar but a spin period $< 3$ ms) may resemble GRB of very long
duration more than conventional radio pulsars.

An unknown amount of baryonic matter will be entrained by the outflowing
pair plasma.  The degree of entrainment and the pair density will vary
across the surface of the disc, both in space and in time, as the magnetic
field and flow geometry vary.  There is abundant opportunity for radiation
by internal shocks as these flows of varying Lorentz factors overtake each
other and interact, as is required to explain the complex time structure
of GRB.

The deposition of a super-Eddington power (for example, by viscous 
dissipation or neutrino transport) within the deep interior of the dense,
massive disc need not imply a strong mass outflow.  Hence pair plasmas
with low baryon loading and high Lorentz factors are possible.  The
Eddington limit only constrains the rate at which radiation can diffuse
through matter.  A ``super-Eddington'' diffusive photon luminosity does not
produce an outflowing wind; it does not occur in a hydrostatic 
configuration.  If a power greater than the Eddington limit (but less than
the binding energy divided by the hydrostatic relaxation time) is injected
into opaque matter it will produce only a quasi-static structural 
relaxation; see \cite{K96} for a recent discussion.

The magnetic field inside the disc produces a stress which leads to an 
outward flow of angular momentum, and inward flow of mass and the release
of gravitational energy, as in the usual theory of accretion discs.  This
process is usually described as dissipation by magnetic viscosity, and its
detailed mechanism, as well as the process by which the energy released is
thermalized, is controversial.  The relevant component of magnetic stress 
(in a thin disc) is $B_r B_\phi / 4 \pi$.  Regardless of the detailed
mechanism of field production, no other magnetic stress is available to
transport angular momentum outward.  A Newtonian estimate for the power 
released by magnetic viscosity is then $$P_{visc} \approx {B_r B_\phi \over 2} \Omega r^2 h, \eqno(3)$$
where $h$ is the disc half-thickness.    

Comparing Equations (1) and (3) and assuming that the $B^2$ on the surface 
of the disc used in Equation (1) is comparable to $B_r B_\phi$ leads to
$${P_{rw} \over P_{visc}} \sim {r \over h} \left({\Omega r \over c}
\right)^3. \eqno(4)$$
The ratio $r/h$ is variously estimated at 1--10; $(\Omega r / c)^3 \sim
0.1$ near a neutron star's surface or a black hole's last stable circular
orbit, but declines $\propto r^{-3/2}$ at larger radii.  Hence near the
surface (or last stable circular orbit) the electrodynamic efficiency
$$\epsilon \equiv {P_{rw} \over P_{rw} + P_{visc}} \eqno(5)$$
is in the range 0.1--0.5, while it is small in the outer regions of a 
disc.  The efficiency of production of nonthermal power (and ultimately
radiation) does not depend on the magnitude of the magnetic field, 
although it does depend on the field geometry.

The energy available from the accretion of $0.1 M_\odot$ is $\sim 10^{53}$
ergs, assuming maximally rotating Kerr geometry, so that at least $\sim 
10^{52}$ ergs are available for GRB emission.  This is sufficient to 
explain the energetics of GRB at cosmological distances so long as the
efficiency of emission by the relativistic shocks is not small. 
That portion of the accretional energy which does not appear in
the relativistic wind is thermalized within the disc and either emitted as
neutrinos or swept into the black hole (or onto the rotating neutron star)
along with the accreted matter.
\section{Magnetic Fields}
Can magnetic fields of $10^{15}$ gauss be justified?  When two neutron
stars coalesce they quickly (in a time $\sim (R^3 / GM)^{1/2} \sim 10^{-4}$
s) settle down to an axisymmetric but differentially rotating
configuration.  The near-adiabaticity of neutron star coalescence implies 
that thermal convection is weak.  Because the flow is axisymmetric the 
Cowling theorem (\cite{S92})
establishes that it cannot generate a magnetic field by a dynamo process.
It is possible that a weak dynamo may occur because of small deviations
from exact axisymmetry, but these cannot be estimated.

Differential rotation generates azimuthal field from radial field,
requiring only that the field be frozen into the matter, according to
the equation
$${dB_\phi \over dt} = B_r {d\Omega \over d\ln r} \approx - {3 \over 2}
B_r \Omega. \eqno(6)$$
This is not a dynamo, as the radial field is not regenerated, and will
eventually decay resistively.

Ropes of (nearly azimuthal) magnetic flux are buoyant, and will rise to
the surface of the disc on a time scale
$$t_b \sim {h \over v_A} \sim {h (4 \pi \rho)^{1/2} \over B}, \eqno(7)$$
where $v_A = B/(4 \pi \rho)^{1/2}$ is the Alfv\'en speed and $\rho$ a mean
density (depending on the geometry, this may involve the Parker
instability, but the time scale is determined by dimensional considerations
alone).  A rough (statistically) steady state will be reached when the
rate of field growth (Equation 6) equals its rate of loss by buoyancy
(Equation 7).  Using the likely dominant component $B_\phi$ for $B$ in
Equation (7) permits its magnitude to be estimated.  The result is
$$B_\phi \sim \left[{3 \over 2} B_r \Omega h (4 \pi \rho)^{1/2}
\right]^{1/2} \sim 10^{15}\,B_{r13}^{1/2} \Omega_4^{1/2} h_6^{1/2} 
\rho_{13}^{1/4}\ {\rm gauss}, \eqno(8)$$
where $B_{r13} \equiv B_r /(10^{13}\,{\rm gauss})$, $\Omega_4 \equiv \Omega
/(10^4\,{\rm s}^{-1})$, $h_6 \equiv h /(10^6\,{\rm cm})$ and $\rho_{13}
\equiv \rho / (10^{13}\,{\rm g}\,{\rm cm}^{-3})$.

A plausible origin of $B_r$ is the pre-coalescence magnetic fields of the
neutron stars.  Values as large as $\sim 10^{13}$ gauss are observed for 
some radio pulsars, justifying the use of $B \sim 10^{15}$ gauss if the 
Parker instability or buoyant rise convert $B_\phi$ to $B_z$ with 
reasonable efficiency.  The relative magnitudes of the three components 
$B_r$, $B_\phi$ and $B_z$ are uncertain, which introduces similar
uncertainties into estimates such as Eq. (4).  Millisecond pulsars are
observed with fields as small as $\sim 10^9$ gauss (other neutron stars may 
have yet smaller fields but be unobservable for that reason), implying
$B \sim 10^{13}$ gauss.
\section{Time Scales}
\subsection{Duration}
The magnetic fields determine the rate of accretion and the rate of 
emission of GRB power.  If $B_z$ is comparable to the $B_\phi$ estimated 
from Eq. (8) then
$$P_{rw} \sim 10^{51}\,B_{r13} \Omega_4^5 h_6 \rho_{13}^{1/2} r_6^6\ 
{\rm erg/s}, \eqno(9)$$
where $r_6 \equiv r/(1.5 \times 10^6\,{\rm cm})$.
A total GRB energy $E \sim 10^{51}$ ergs implies a duration, assuming
efficient conversion of relativistic wind energy, of
$$t_d \sim 1\,B_{r13}^{-1} \Omega_4^{-5} h_6^{-1} \rho_{13}^{-1/2} r_6^{-6}\ 
{\rm s}. \eqno(10)$$

Short GRB may be explained by the coalescence of neutron stars with
magnetic fields roughly comparable to those observed for most radio
pulsars.  The shortest GRB ($t_d \sim 10^{-2}$ s) may require somewhat 
larger fields (not implausible, because pulsars with such large fields would
spin down rapidly and have short observable lives), or somewhat different
values of other parameters, or may instead be explained by smaller values of
$E$ and $P_{rw}$, consistent with our poor quantitative understanding of GRB
energetics.  

GRB with durations as long as $\sim 10^4$ s may be explained by the 
coalescence of millisecond pulsars with $B_r \sim 10^9$ gauss, and even 
longer durations (such as the $\sim 10^5$ s required to explain the ``Gang 
of Four'' apparent repetitions of October 27--29, 1996 as a single event;
\cite{Me96,C97}) are possible.  In the present model long durations pose no
intrinsic difficulty, and need not be associated with unusually soft 
spectra, in contrast to external shock models in which they imply low 
Lorentz factors and low radiative efficiency.  Very long GRB are faint, on 
average, in any model in which the total energy of a GRB is limited, and 
are therefore difficult to detect; they may be more frequent than is 
apparent from intensity (or rate-of-rise) selected samples.
\subsection{Time scales and substructure}
In the present model the magnetic field rearranges itself on the time
scale $t_b$, which may be rewritten
$$t_b \sim (4 \pi \rho)^{1/4} \left({2 h \over 3 B_r \Omega} \right)^{1/2} 
\sim 0.01\ h_6^{1/2} \rho_{13}^{1/4} B_{r13}^{-1/2} \Omega_4^{-1/2}\ 
{\rm s}. \eqno(11)$$
Equivalently, $B_r$ may be eliminated in favor of the duration $t_d
\equiv E/P_{rw}$, yielding
$$t_b \sim \left({t_d \over 2 E}\right)^{1/2} (4 \pi \rho)^{1/2} h^{3/2}
r \Omega^{1/2} \sim 0.04\ \left({10^{51}\,{\rm erg}\,{\rm s}^{-1} \over 
P_{rw}} \right)^{1/2} \rho_{13}^{1/2} h_6^{3/2} r_6 \Omega_4^{1/2}\ {\rm s}.
\eqno(12)$$
The model predicts that the GRB intensity does not vary much on time scales
shorter than $t_b$.  This appears to be consistent with the data.
\subsection{Complexity}
The complex time structure of GRB is notorious.  It led \cite{SF73}
to point out a qualitative resemblance to the time structure of
Solar flares.  Many different qualitative forms are seen: continuous 
irregular fluctuations, isolated peaks separated by longer intervals with
no detected emission, single simple peaks, peaks with numerous 
sub-peaks$\ldots$.  It is beyond the capability of any theoretical model to 
predict this complex structure; it is a formidable task even to construct 
suitable measures to describe it statistically.  

In the present model the field rearranges itself on the time scale $t_b$,
and there are $t_d/t_b \sim 10^2$--$10^3$ (the numerical value depending on
poorly known parameters) rearrangements in a GRB of 10 s duration;
note that $t_d$ is the length of time during which the GRB radiates
strongly, and may be much less than the measured pulse length if there
are peaks of intensity separated by periods of much lower or zero emission.
This permits a great variety of complex time structures.  Their details 
depend on unknown details of the disc's magnetohydrodynamics.  For 
comparison, consider the problem of explaining the Solar magnetic cycle.  
This is incompletely understood, even with the aid of a great body of data.
Its observed complexity argues for the plausibility of obtaining the variety
of GRB time structure as the result of disc magnetohydrodynamics.

The preceding is only an argument for the plausibility of obtaining the
observed GRB temporal phenomenology.  It is not a demonstration that
the observed heterogeneity and complexity can be obtained.  Nevertheless,
the situation is better than for external shock models, which \cite{SP97a}
showed could not explain the data.  Even before their 
argument was made, external shock models depended on the hopeful wish that 
the temporal complexity of GRB could be attributed to the (also poorly
understood) spatial complexity of the interstellar medium.

The failure to demonstrate that the variety of observed pulse forms must
occur might be used as an argument against the present model.  If it is 
accepted that all classical GRB are produced by a single kind of event then
this heterogeneity could be used equally as an argument against any model, 
because we have never understood any physical process which produces complex
and heterogeneous pulse forms well enough to predict these forms.  The 
alternative hypothesis that many different kinds of events produce GRB is 
even more unlikely: apart from the observed homogeneity among GRB in energy 
scale, spatial distribution, spectral properties and even the fact of 
complex pulse structure, this hypothesis requires the construction of 
several satisfactory models, when it is hard enough to find even one!  All 
that can be asked is that a model {\it permit} temporal complexity and 
heterogeneity.
\subsection{Quasiperiodic oscillations}
The magnetic field of the accretion disc will not, in general, be
axisymmetric.  If the deviation from axisymmetry is large the resulting model
resembles magnetic dipole emission (though of a relativistic wind rather
than of an electromagnetic wave) more closely than an axisymmetric rotator;
Equations (1) and (2) remain valid in either case.  The particle flux and 
the observed radiation may therefore be modulated at the rotation frequency,
which varies with $r$ in the disc.  This time dependence will be convolved 
with the time dependence of acceleration and radiation of the radiating 
particles and geometric delays arising from a range of signal path lengths 
(\cite{R75,K94}), and may thereby be unobservable.  However, the 
acceleration and radiation times may be short (\cite{K94,SNP96,SP97b}), and 
the rotational modulation should be looked for in gamma-ray data of 
sufficiently high temporal resolution.  It is also conceivable, though
probably unlikely, that sufficient coherent emission occurs to produce a
detectable signal at radio frequencies, as in radio pulsars.

The expected signal would consist of quasi-periodic oscillations (QPO).  
There is predicted to be a high frequency cutoff at the maximum rotational
frequency of the disc.  For a disc surrounding a maximally rotating Kerr
black hole of 2.8 $M_\odot$ (the expected result of the coalescence of two
neutron stars) the frequency in the source frame of the last stable circular
orbit (\cite{ST83}) is 5770 Hz.  If QPOs
with an upper frequency cutoff below this value are observed, cosmological 
distances will be verified and the redshifts of individual GRB measured.  
The maximal rotational frequency of a differentially rotating neutron star
depends on its distribution of angular momentum and is significantly
smaller.  It might be calculable from simulations of neutron star
coalescence, but the interpretation of the data would be more ambiguous.

The spectral power density will drop off towards lower frequencies 
because both the energy release and the relativistic wind efficiency 
(Equation 5) decline with $r$ (a naive estimate is that the power density 
varies $\propto \Omega^{2/3}$).  Discs resulting from neutron star
coalescence may have an abrupt outer edge, because viscosity has not had
time to extend them to large radii or because exothermic nuclear reactions
in expanded neutronized material expel their lower density regions, so that
there may also be an abrupt low frequency cutoff.  The distribution of power
will give some indication of the radial structure and distribution of
magnetic fields in the disc.
\section{Magnetic Reconnection}
It was natural (\cite{SF73,K82,NPP92,K94}) to consider magnetic
field reconnection as the explanation of the complex time structure of GRB.
This is unlikely.  It would require that the entire GRB energy (divided by
an efficiency $< 1$) pass through the form of magnetic energy.  This would
require a strong dynamo, but we expect any dynamo to be weak (if present
at all) because the flow is axisymmetric, at least to a first approximation.

The mechanism proposed in this paper has the advantage (in comparison to
magnetic reconnection) that the same magnetic flux is a source of emitted
power for an indefinite time.  Even a small (compared to the accretion 
energy) magnetic energy can be the conduit through which the entire GRB
energy flows (because a GRB is very long compared to $r/c$).  In contrast,
in a magnetic annihilation model the magnetic energy must be regenerated
rapidly by a dynamo, and all the radiated energy must have at one time 
assumed the form of magnetostatic energy.

The relativistic wind has a power (Equation 2) which is the magnetic
energy multiplied by a speed $O(r^3 \Omega^3 /c^2) \sim O(c/10)$.  In 
order to be equally powerful, magnetic annihilation would have to act on a
surface covering most of the inner accretion disc, throughout the duration
of a GRB, with a reconnection speed $v_f \sim c/10$.  This is surely
optimistic.  In general, $v_f$ must be less than $v_A$, which is small in
dense matter (\S3) although it may be relativistic where the wind is 
generated.

Any process which produces the observed power density produces an
equilibrium pair plasma, as discussed in \S2.  The electrical conductivity
of such a plasma
$$\sigma \sim {m_e c^3 \over e^2} \sim 10^{23}\ {\rm s}^{-1}, \eqno(13)$$
as may be found by dimensional analysis or by estimating the number of
charge carriers and their scatterers (charged particles and photons).  The
conductivity is approximately independent of temperature for $k_B T > m_e
c^2$ because the densities of carriers and of scatterers each vary
$\propto T^3$.  The relativistic temperature makes counterstreaming plasma
instabilities unlikely, and their high density makes the relative drift
velocity of electrons and positrons small.

In a reconnecting current sheet of thickness $\ell$ the time $\ell / v_f$
required for the flow to regenerate the magnetic flux may be equated to the
resistive dissipation time $\sigma \ell^2 / c^2$.  The result is
$$\ell \sim {c^2 \over v_f \sigma} \sim {c \over v_f} {e^2 \over m_e c^2}
\sim {c \over v_f} r_e, \eqno(14)$$
where $r_e \equiv e^2 /{m_e c^2} = 2.82 \times 10^{-13}$ cm is the classical
electron radius.  The electron density at $k_B T \sim m_e c^2$ is $n_e \sim
m_e^3 c^3 / \hbar^3 \sim 10^{31}$ cm$^{-3}$.  The discreteness of the
charge carriers limits $\ell > n_e^{-1/3} \sim \hbar / (m_e c) \sim 4 \times
10^{-11}$ cm.  This is consistent with Equation (14) only if
$${v_f \over c} < {e^2 \over \hbar c} \approx {1 \over 137}. \eqno(15)$$
This is a general limit on the speed of magnetic reconnection in a
relativistic pair plasma.  In the inner region of a disc around a black hole
or a neutron star magnetic reconnection is, at best, an order of
magnitude less powerful than the relativistic wind flowing on open field
lines.  Magnetic flux is probably destroyed by accretion onto the black
hole or advection to infinity in the wind, rather than by reconnection.
\section{Unified Model of GRB, AGN and BHXS}
These three classes of objects have certain qualitative similarities
(assuming the present model of GRB), despite large quantitative differences
in their luminosities, masses, and time scales.  They all are powered by
accretion onto a central black hole.  They all produce nonthermal radiation.
They all show evidence for relativistic beaming and relativistic bulk 
motion.  They all fluctuate irregularly in intensity.  These similarities 
suggest that it may be possible to construct a single unified model for them
all.  The present model for GRB resembles the \cite{B76,L76} model for AGN, 
although they assumed (probably unnecessarily) that the magnetic dipole 
moment was aligned with the rotational axis.  These models may be scaled to 
stellar mass BHXS such as Cyg X-1.  The most important difference is that in
GRB the relativistic wind is thermalized to an equilibrium pair plasma,
while at the lower power densities of AGN and BHXS it remains transparent
and nonequilibrium.

I consider a class of models of AGN and BHXS in which the electromagnetic
energy is converted to the energy of accelerated particles close to the black
hole.  Very energetic particles take the place of the equilibrium pair
plasma in GRB.  In another class of models of AGN and BHXS the disc radiates
vacuum electromagnetic waves instead of energetic particles.  At much
greater radii these waves accelerate particles, just as in GRB the pair 
plasma accelerates particles in distant shocks.  These two classes of models
can be comparably efficient particle accelerators.  I do not consider the
vacuum wave model further because it is less closely analogous to the GRB
model (which cannot be a vacuum wave model because the energy density leads
to creation of an equilibrium pair plasma), and because external plasma
injection or pair breakdown (\S6.7) are likely to fill the wave zone with
energetic particles
\subsection{Nonthermal efficiency}
All three classes of objects show a significant amount of nonthermal 
emission.  In GRB all the observed emission appears to be nonthermal, 
although it is not the primary radiation emitted by the central engine but 
rather the consequence of a shock produced by or in the relativistic 
outflow; the high energy density at the source thermalizes the relativistic 
wind.  Still, the wind is produced by a fundamentally nonthermal process.  In 
many AGN a substantial fraction of the power appears as nonthermal visible
synchrotron radiation or high energy gamma-rays.  Extragalactic radio 
sources appear to be the consequence of the acceleration of relativistic 
particles in AGN.  The case for nonthermal emission in BHXS is plausible
but less compelling.  It is a natural explanation of the complex 
multi-component spectra observed for the best studied example (Cyg X-1).  
The superluminal radio components and jets observed in some of these objects
certainly require acceleration of relativistic particles.  Cyg X-3 also 
shows strong outbursts of nonthermal radio emission, although there is no 
direct evidence it contains a black hole.

In GRB the efficiency $\epsilon$ (Equation 5) is not directly measured
because virtually all the thermal radiation emerges as neutrinos and is 
essentially undetectable.  For the lower density accretion flows of AGN
and BHXS neutrino emission is negligible, and the thermal radiation produced
by viscous heating is directly observable.  In \S2 I argued that $\epsilon 
\sim 0.1$--0.5 is likely, independent of the magnitude of the magnetic field
(but depending on its unknown orientation and spatial structure).  This is 
consistent with observations of AGN; the likelihood of relativistic beaming 
precludes quantitative comparisons.  This range of $\epsilon$ is also 
consistent with observations of Cyg X-1 and other BHXS if the harder 
components of their spectra are either nonthermal or the thermal emission of
optically thin matter heated by nonthermal particles.

The measured nonthermal efficiency is also affected by radiation trapping
(\cite{K77}); if the mass accretion rate exceeds the nominal Eddington rate
the excess mass is swallowed by the black hole, but the emergent luminosity
(in thermal radiation which diffuses through the accretion flow) is
limited to slightly less than $L_E$ (\cite{ECK88}).  An analogous limit 
applies to the unobserved neutrino luminosity.  In AGN and BHXS this can 
lead to an apparent nonthermal efficiency $\epsilon \to 1$ and $P_{rw} \gg
L_{th}$, because the nonthermal wind luminosity is proportional to the 
accretion rate and is not subject to the Eddington limit, while the emergent
thermal luminosity $L_{th}$ cannot exceed $L_E$ even if $P_{visc} \gg L_E$.
\subsection{Time structure}
Substitution of elementary estimates of steady nonrelativistic accretion
discs (the GRB disc may be steady on the fastest time scale $r/c$, even
though it is unsteady on the timescale $t_b$) into Equation (12) yields
$$t_b \sim {r \over c (\alpha \epsilon_a)^{1/2}}, \eqno(16)$$
where $\alpha$ is the conventional ratio of viscous stress to pressure
and $\epsilon_a c^2$ is the energy per unit mass released by accretion.
For $\alpha = 0.1$ (much larger than implied for GRB) and
$\epsilon_a = 0.06$ (appropriate to the last stable circular orbit around
a Schwarzschild black hole)
$$t_b \sim 4 \times 10^{-4} {M \over M_\odot}\ {\rm s}. \eqno(17)$$
This estimate is roughly an order of magnitude longer, but has the same
scaling, as the most naive estimate of $\sim r/c$.  For a maximally
rotating Kerr black hole the corresponding estimate is
$$t_b \sim 2 \times 10^{-5} {M \over M_\odot}\ {\rm s}, \eqno(18)$$
comparable to the naive estimate.  

In comparing to AGN it should be remembered that the nonthermal power is not
Eddington limited, nor is the thermal radiation of unbound gas clouds heated
by the nonthermal power; the black hole's mass therefore cannot be estimated
from the luminosity, but only from its direct gravitational influence on 
surrounding matter.  The matter in such clouds is driven away by radiation
pressure, and must be resupplied, for example by the infall and disruption
of stars (the infall of opaque stars is an illustration of the principle
that there is no Eddington bound on the rate of mass accretion).

The complex multi-state behavior of Cyg X-1 may qualitatively be 
attributed to changes in the magnetic geometry, which affect $\epsilon$,
but the magnetic cycles of accretion discs are too poorly understood
to admit a more quantitative understanding.  There is some resemblance
between the irregular multi-peaked structure of GRB and the appearance and 
disappearance of hard components in the spectrum of Cyg X-1.  Even in a 
thermal model, these hard components require the presence of matter much 
hotter than black body equilibrium temperatures (which are $< 1$ KeV), and 
this matter may be heated by the nonthermal particles.

It is interesting to note a qualitative similarity between BHXS and GRB.
In one BHXS (Cyg X-1; \cite{W78}) a non-zero time skewness was measured from
an X-ray time series.  Time skewness of the same sense is obvious in many 
GRB, where it is often called the FRED (fast rise, exponential decay) pulse 
shape.  Searches for time skewness in AGN time series have so far been 
unsuccessful (\cite{PR97}), but do not exclude it.
\subsection{Particle acceleration}
Pulsars and GRB (in the present model) have large electric fields which
lead to pair production.  A supermassive black hole in a galactic nucleus, 
or a stellar mass black hole in a mass transfer binary, is surrounded by a 
complex accretional gas flow.  Significant sources of mass include the 
companion star (in the BHXS), the galactic interstellar medium (in the AGN) 
and the surface of the outer parts of the accretion disc.  Although the flow
is not understood in detail, it is plausible that some of this gas has 
sufficiently little angular momentum (or loses its angular momentum at large
radii) to permit accretion on the axis of rotation, and can fill all 
directions around the black hole (as was found in the calculations of 
\cite{ECK88}).  A pulsar-like vacuum is not likely.  The space charge 
density required to neutralize the corotational electric field (\cite{GJ69})
is small, and may readily be supplied by this plasma.  Pair production is 
therefore not required for the extraction of energy from the rotating disc
in AGN and BHXS, although it may occur.

Near a luminous object accelerated electrons and positrons would be rapidly
slowed by Compton scattering on the thermal radiation field 
(\cite{J65,KS74}); the energy loss length $\ell_C$ of an electron with
Lorentz factor $\gamma$ in an isotropic radiation field of intensity $L_{th}
/ (4 \pi r^2)$ near a mass $M$, assuming the Thomson cross-section, is
$$\ell_C = {m_e \over m_p} {1 \over \gamma} {L_E \over L_{th}} {r c^2 \over 
GM} r, \eqno(19)$$
The first two factors are each $\ll 1$, and the last two are not much 
greater than unity in the inner disc of a luminous object, so that $\ell_C 
\ll r$.  This result is also approximately valid for anisotropic radiation 
fields, except in the extreme case of a particle moving accurately in the 
direction of a narrowly collimated beam of radiation.

Given a value for $B$, it is possible to calculate the maximum energy an
electron achieves, and to estimate its radiation.  Equating the magnetic 
stress to that required to supply a bolometric luminosity $L_b$ (including 
relativistic wind, thermal radiation and radiation advected into the black
hole) yields
$${B^2 \over 8 \pi} \sim {1 \over 2} {L_b \over L_E} \left({GM \over r c^2}
\right)^{3/2} {r \over h} {c^4 \over GM \kappa} \sim 3 \times 10^7\ {L_b 
\over L_E} \left({10 GM \over r c^2}\right)^{3/2} {r \over 10 h} {10^8 
M_\odot \over M}\ {{\rm erg} \over {\rm cm}^3},  \eqno(20)$$
where $\kappa$ is the Thomson scattering opacity.  This estimate assumes 
only that the viscosity is magnetic and that $B^2 \sim \langle B_r B_\phi 
\rangle$; it is not dependent on an assumption of equipartition or on any 
theory of $\alpha$, and is derived from the accretion rate implied by $L_b$
alone.

Equating the energy gained in a length $\ell_C$ to the energy $m_e c^2 
\gamma$ lost by Compton scattering, using Equation (19), yields the 
maximum Lorentz factor 
$$\gamma_C \sim \left({eEr \over m_p c^2} {r c^2 \over GM} {L_E \over 
L_{th}} \right)^{1/2}. \eqno(21)$$
Using $E \sim v B/c \sim B (GM/r c^2)^{1/2}$ and Equation (20) yields
\setcounter{equation}{21}\begin{eqnarray}
\gamma_C &\sim &\left({r c^2 \over GM}\right)^{1/8} \left({L_{th} \over 
L_E}\right)^{-1/2} \left({r \over h}\right)^{1/4} \left({GNm_e^2 \over e^2}
\right)^{1/4} \left({L_b \over L_E}\right)^{1/4} \nonumber \\ &\sim &1.0 
\times 10^4 \left({M \over M_\odot}\right)^{1/4} \left({r c^2 \over 10 GM}
\right)^{1/8} \left({L_{th} \over L_E}\right)^{-1/2} \left({r \over 10 h}
\right)^{1/4} \left({L_b \over L_E}\right)^{1/4}, \end{eqnarray}
where $N \approx 1.2 \times 10^{57}\,M/M_\odot$ is the ratio of the black
hole mass to the proton mass.  It is amusing to express $\gamma_C$ in terms
of fundamental constants, dropping factors of order unity which depend on
the properties of the individual object, and noting that for a Chandrasekhar
mass $M_{Ch}$ $N \approx (\hbar c/G m_p^2)^{3/2}$:
$$\gamma_C \sim \left({m_e \over m_p}\right)^{1/2} \left({e^2 \over \hbar c}
\right)^{-1/4} \left({G m_p^2 \over \hbar c}\right)^{-1/8} \left({M \over 
M_{Ch}}\right)^{1/4}.  \eqno(23)$$
\subsection{Compton gamma-rays}
Equation (22) directly gives an upper bound on the Compton scattered photon
energy $\gamma_C m_e c^2$, which is $\sim 10^{12}$ eV for typical AGN
parameters and $\sim 10^{10}$ eV for typical BHXS parameters.  
This explains, at least qualitatively, the production of TeV gamma-rays in 
AGN such as Mrk 421 (\cite{G96}) and Mrk 501 (\cite{Q96}).  

The actual spectral cutoff depends on the spectrum of thermal photons; if 
their energy $\hbar \omega_{th} \sim m_e c^2 / \gamma_C$ the cutoff will be 
$\sim \gamma_C m_e c^2$; if $\hbar \omega_{th} < m_e c^2 / \gamma_C$ the 
cutoff will be $\sim \gamma_C^2 \hbar \omega_{th}$; if $\hbar \omega_{th} > 
m_e c^2 / \gamma_C$ the cutoff is $\sim \gamma_C m_e c^2$ but Equation (22) 
then underestimates $\gamma_C$ because of the reduction in Klein-Nishina 
cross-section and the discreteness of the energy loss.  The observed spectra
of AGN and BHXS are so complicated (it is also unclear which components are 
emitted at the small radii at which electron acceleration is assumed to 
occur) that it is difficult to be quantitative.  Visible radiation from AGN 
and soft X-rays from BHXS place Compton scattering marginally in the 
Klein-Nishina range (note that both the black body $\hbar \omega_{th}$ and 
$m_e c^2 / \gamma_C$ scale $\propto M^{-1/4}$), but quantitative
estimates depend on the unknown factors in parentheses in Equation (22).

Because the Compton scattering power of an electron and the frequency of the
scattered photon each are proportional to $\gamma^2$, while the rate of 
energy gain in an electric field is independent of $\gamma$, there is 
expected to be a broad peak in $\nu F_\nu$ around the cutoff frequency, with
$F_\nu \propto \nu^{1/2}$ at lower frequencies (the same slope as for 
synchrotron radiation, for the same reasons).  In principle, the weak 
dependence of $\gamma_C$ on $M$, $L_{th}$ and $L_b$ in Equation (22) could 
be tested if $M$ were estimated independently of the luminosities, for 
example, from the time scale of variations.

The majority of the power which goes into electron acceleration may appear
as Compton scattered gamma-rays of energy $\sim \gamma_C m_e c^2$.  This
is half $P_{rw}$ in a proton-electron wind and all of $P_{rw}$ if pairs
are accelerated.  As discussed in \S6.1 this can far exceed $L_{th}$ if
the disc is undergoing highly supercritical accretion.  This may explain
the dominance of the emitted power by energetic gamma-rays in some AGN.
Supercritical accretion by black holes of comparatively low mass also
permits more rapid variability than accretion at the Eddington limit.
\subsection{Synchrotron radiation}
Electrons with Lorentz factors up to that given by Equation (22) may radiate
synchrotron radiation, assuming an isotropic distribution of pitch angles,
at frequencies up to
$$\nu_{synch} \sim \left({3 \over 8 \pi^2}\right)^{1/2} \left({GM \over r 
c^2} \right)^{1/2} {L_b \over L_{th}} {r \over h} {m_e c^3 \over e^2} \sim 7 
\times 10^{22}\,{L_b \over L_{th}} \left({10 GM \over r c^2}\right)^{1/2}
{r \over 10 h}\ {\rm Hz}.  \eqno(24)$$
The predicted spectral index for $\nu < \nu_{synch}$ is $-1/2$, because a
uniform accelerating electric field produces a particle distribution 
function with energy exponent zero.

Unlike the Compton cutoff, the synchrotron cutoff is independent of the
mass of the black hole.  Because the magnetic and thermal photon energy
densities may be comparable, synchrotron radiation may be energetically
important.  The ratio of synchrotron to Compton scattering powers, assuming
an isotropic electron distribution, is the ratio of the magnetic to the
thermal energy densities $U_B/U_{th}$:
$${P_{synch} \over P_{Compt}} = {U_B \over U_{th}} \sim {1 \over 2} {L_b
\over L_{th}} {r \over h} \left({GM \over r c^2}\right)^{-1/2}. \eqno(25)$$

Equations (24) and (25) suggest the possibility of synchrotron radiation 
with significant power up to $\sim$ GeV energies.  In most circumstances 
this is probably a great overestimate because electrons may be effectively 
accelerated only parallel to the magnetic field---a component of $E$ 
perpendicular to $B$ does not effectively accelerate charged particles 
unless it varies at their cyclotron frequency, unlike the nearly steady 
corotational electric field.  

Internal shocks in the relativistic wind may be as essential to radiation in
AGN as in GRB, for in them plasma turbulence may partially isotropize the 
electron distribution, making effective synchrotron radiation possible.  If 
the pitch angles remain small the frequency of synchrotron radiation is 
reduced and it is emitted nearly parallel to the direction of the electrons'
motion, the magnetic field, and the Compton scattered gamma-rays.  This can 
be described as relativistic bulk motion of the electrons and their 
associated radiation field, and may be necessary to avoid absorption of the 
photons by $\gamma-\gamma$ pair production.
\subsection{Curvature radiation}
The electrons also radiate curvature radiation (on the magnetic field lines
of radii of curvature $\sim r$ at frequencies up to
\setcounter{equation}{25}\begin{eqnarray}
\nu_{curv} &\sim &{1 \over 2 \pi} \left({GM \over r c^2}\right)^{5/8} \left(
{L_{th} \over L_E}\right)^{-3/2} \left({L_b \over L_E}\right)^{3/4} \left({r
\over h}\right)^{3/4} {m_e^{3/2} c^3 \over m_p^{3/4} e^{3/2} (GM)^{1/4}}
\nonumber \\
&\sim &3 \times 10^{15} \left({10 GM \over r c^2}\right)^{5/8} \left({L_{th}
\over L_E}\right)^{-3/2} \left({L_b \over L_E}\right)^{3/4} \left({r \over
10 h}\right)^{3/4} \left({M_\odot \over M}\right)^{1/4}\ {\rm Hz}. 
\end{eqnarray}
Curvature radiation is insignificant, and its power is small.  Unlike in
pulsars, it cannot cause pair production because the electron Lorentz factor
is limited by Compton scattering.
\subsection{Pair production}
The most energetic gamma-rays of energy $E_\gamma$ may produce 
electron-positron pairs by interacting with the thermal photons.  The 
condition for this to occur (assuming an isotropic thermal radiation field)
$$E_\gamma \hbar \omega_{th} \sim \gamma_C m_e c^2 \hbar \omega_{th} > 
(m_e c^2)^2, \eqno(27)$$
is equivalent to the condition for the breakdown of the Thomson 
approximation to Compton scattering.  It appears to be met in AGN and BHXS, 
taking the observed thermal spectra and assuming all factors in parentheses 
in Equation (22), except that involving the mass, are $O(1)$.  If the 
thermal radiation field is assumed to be a black body then Equation (27)
can be rewritten
$$\left({GM \over r c^2}\right)^{3/8} \left({r \over h}\right)^{1/4} \left(
{L_b \over L_{th}}\right)^{1/4} \left({e^2 \over \hbar c}\right)^{-3/4} > 1.
\eqno(28)$$  It is clear that this condition is generally met; if (as is 
likely) the thermal spectrum is harder than a black body at the effective 
temperature the inequality holds even more strongly.  Pair production by
interaction between gamma-rays produced by Compton scattering of the
accelerated electrons and thermal photons takes the place of pair
production by curvature radiation which occurs in pulsars.  
\subsection{Proton acceleration}
If there were no pair production, the accelerated plasma would consist of 
protons (and nuclei) and electrons.  Even in the presence of pair 
production, protons may be accelerated along with the positrons.  This is 
important in intense sources of thermal radiation, such as BHXS and AGN, 
because proton-photon scattering is negligible below the pion production 
threshold.  Even above threshold, the effective energy loss cross-section 
(\cite{G66}) is a fraction $f \sim 10^{-4}$ of the Thomson cross-section.  
This permits accretion discs around black holes in AGN and BHXS to be 
efficient proton accelerators (as has previously been discussed by 
\cite{LB69,KE86,K91} in other
models).  Very high energy gamma-rays may then result from photoproduction of 
$\pi^0$ on thermal radiation.  Collisions of protons with nucleons also
produce high energy radiation directly from $\pi^0$ and indirectly by 
Compton scattering of the very energetic $e^\pm$ produced by $\pi^\pm \to 
\mu^\pm \to e^\pm$ (\cite{K91,DL97}).

The energy loss length of a proton is, in analogy to Equation (19),
$$\ell_p \approx {1 \over \gamma f} {L_E \over L_{th}} {r c^2 \over GM} r.
\eqno(29)$$
Unlike Thomson scattering, this process has an energy threshold because of 
the $\pi^0$ rest mass of 135 MeV.  If the thermal spectrum consists of 
visible light, the acceleration of protons is not restrained until $\gamma_p
\sim 10^8$.  The result analogous to Equation (22) is
\setcounter{equation}{29}\begin{eqnarray}
\gamma_p &\sim &\left({r c^2 \over GM}\right)^{1/8} \left({L_{th} \over L_E}
\right)^{-1/2} \left({r \over h}\right)^{1/4} \left({GNm_e^2 \over e^2}
\right)^{1/4} \left({L_b \over L_E}\right)^{1/4} f^{-1/2} 
\nonumber \\ &\sim &1.0 \times
10^6\ \left({M \over M_\odot}\right)^{1/4} \left({r c^2 \over 10 GM}
\right)^{1/8} \left({L_{th} \over L_E}\right)^{-1/2} \left({r \over 10 h}
\right)^{1/4} \left({L_b \over L_E}\right)^{1/4}. \end{eqnarray}
For an AGN with $M \sim 10^8 M_\odot$ this yields a limiting $\gamma_p \sim 
10^8$, approximately the threshold at which $\pi^0$ photoproduction begins.
As for pair production, $\gamma_p$ and the thermal photon energy scale 
reciprocally with $M$, so that Equation (30) is (barely) applicable at all
$M$.

The accelerated protons are not energetic enough to explain the highest
energy cosmic rays, even though the nominal potential drop across the
disc may approach $\sim 10^{20}$ eV for discs around supermassive black
holes.  The possible $\sim 10^{17}$ eV protons in AGN would produce $\sim
10^{16}$ eV photons, but these are not observable at great distances because
of pair production on the microwave background radiation.  In BHXS the
corresponding photon energy is $\sim 10^{14}$ eV.
\subsection{Quasiperiodic oscillations}
AGN may show QPO, just as suggested for GRB, but with typical periods of
order hours to days, depending on their masses in an elementary way 
(Equations 16--18).  No such QPO have been found in the extensive body of 
visible light data on AGN.  This may be explained if the visible light is
produced far from the central object or by thermal radiation from a disc,
which may be nearly azimuthally symmetric.  It may be more fruitful (though
more difficult observationally) to search for QPO in the energetic
gamma-rays produced as particles are accelerated along magnetic field lines 
closer to the rotating disc, which would be expected to show the greatest 
deviations from axisymmetry, as in pulsars.
\section{Discussion}
The demonstration by \cite{SP97a} that the complex time structure
of GRB must be intrinsic to their energy source has forced the rejection of
most previous models.  I have here outlined a model which may solve this
problem.  It predicts a minimum time scale of variations and suggests that
QPO of predictable frequency may be observable.  Observation of the 
predicted QPO would be strong evidence in favor of the model and would 
(given the assumed Kerr black holes) immediately reveal the redshift.  
Alternatively, if an identification with a galaxy of measurable redshift 
were made, a known function of the mass and angular momentum of the black 
hole would be determined, constraining theories of neutron star coalescence.

The development of the GRB model suggests that they result from the same
process, disc accretion onto a black hole, believed to power AGN and
Galactic BHXS.  The parameter regimes are very different, but these
diverse objects may all be the consequence of electrodynamic energy
extraction from accretion discs.  This leads to a unified model which,
by emphasizing their similarities, may help explain all these phenomena.

GRB have a bimodal distribution of durations (\cite{K93}), and long and 
short GRB appear to have different spatial distributions
(\cite{KC96}) and may represent different populations of objects.
The distribution of magnetic fields of neutron stars appears also to be
bimodal (although there are strong observational selection effects which
are difficult to compensate for), which may offer a natural explanation.
One class of pulsars and rotationally modulated accreting neutron stars
has longer periods and $B \sim 10^{12}$ gauss; these may be the origin
of short GRB.  Lower field ($B \sim 10^9$ gauss) neutron stars form a second
class, and are observed as millisecond pulsars and as the presumed low-field
neutron stars in unpulsed low mass X-ray binaries (whose rapid rotation may 
be inferred from their QPO); these may be the origin of long GRB.  High
field neutron stars are produced with large impulses (as inferred
from the space velocities of slow radio pulsars), and low field neutron 
stars (at least, millisecond pulsars found in globular clusters) are
produced with low impulses (\cite{K75}).  This implies that if accurate 
enough coordinates could be obtained (\cite{K97}) short GRB would be found to
be further, on average, from the places of their birth (plausibly the discs
of their host galaxies) than long GRB.  

The coalescence of a neutron star and a black hole may produce a different
event than the coalescence of two neutron stars, but it is not obvious what
the qualitative observable differences would be.  Similarly, the ``silent''
collapse of a rapidly rotating degenerate dwarf may produce a similar
configuration to neutron star coalescence, though of half the mass, but
known degenerate dwarf spins are too slow.
\acknowledgements
I thank T. Piran and R. Sari for discussions, Washington University for the
grant of sabbatical leave, the Hebrew University for hospitality and a
Forchheimer Fellowship, and NASA NAG 52682 and NSF AST 94-16904 for support.

\end{document}